\documentclass{article}

\PassOptionsToPackage{sort, numbers}{natbib}
\usepackage[preprint]{neurips_2023}

\usepackage[utf8]{inputenc} % allow utf-8 input
\usepackage[T1]{fontenc}    % use 8-bit T1 fonts
\usepackage{hyperref}       % hyperlinks

\usepackage{url}            % simple URL typesetting
\usepackage{booktabs}       % professional-quality tables
\usepackage{amsfonts}       % blackboard math symbols
\usepackage{nicefrac}       % compact symbols for 1/2, etc.
\usepackage{microtype}      % microtypography
\usepackage{xcolor}         % colors
\usepackage{natbib}
\usepackage[acronym, hyperfirst=false]{glossaries}
\usepackage{tikz}
\usepackage{graphicx}
\usepackage{subfig}
\usepackage[export]{adjustbox}
\usepackage{bm}
\usepackage{stmaryrd}
\usepackage{musicography}

\usetikzlibrary{shapes, arrows.meta, positioning, shapes.geometric, calc}

\newacronym{ddsp}{DDSP}{Differentiable digital signal processing}
\newacronym{tc}{TC}{timbral consistency}
\newacronym{dac}{DAC}{Descript audio codec}
\newacronym{rvq}{RVQ}{residual vector quantization}
\newacronym{mfccs}{MFCCs}{mel-frequency cepstral coefficients}
\newacronym{midi}{MIDI}{musical instrument digital interface}
\newacronym{clap}{CLAP}{contrastive language-audio pretraining}
\newacronym{fad}{FAD}{Fréchet audio distance}
\newacronym{mad}{MAD$_\mathrm{pitch}$}{median absolute deviation in semitones}
\newacronym{mos}{MOS}{mean opinion scores}

\definecolor{color1}{HTML}{320a5e}
\definecolor{color2}{HTML}{781c6d}
\definecolor{color3}{HTML}{bc3754}
\definecolor{color4}{HTML}{fbb61a}
\definecolor{color5}{HTML}{fcffa4}

\title{InstrumentGen: Generating Sample-Based Musical Instruments From Text}

\author{%
  \And{Shahan Nercessian$^{*}$ and Johannes Imort$^{*}$} \\
  Native Instruments \\%\\
  \texttt{\{firstname.lastname\}@native-instruments.com} \\\\
  {\footnotesize $^{*}$ Equal contribution}
}

\begin{document}
\maketitle

\begin{abstract}
We introduce the \emph{text-to-instrument task}, which aims at generating sample-based musical instruments based on textual prompts. Accordingly, we propose \emph{InstrumentGen}, a model that extends a text-prompted generative audio framework to condition on instrument family, source type, pitch (across an 88-key spectrum), velocity, and a joint text/audio embedding. Furthermore, we present a differentiable loss function to evaluate the intra-instrument timbral consistency of sample-based instruments. Our results establish a foundational text-to-instrument baseline, extending research in the domain of automatic sample-based instrument generation.
\end{abstract}

\section{Introduction}
The synthesis of sounds and corresponding interfaces for controlling their timbre form a seminal topic in audio research \cite{narita_ganstrument_2023}. %Sound synthesis advances in the digital domain have transformed the ways in which music producers realize their artistic visions today. 
Meanwhile, generative models have been successfully applied to images and text, where their convincing ability to draw novel samples from learned data distributions has already proved to be disruptive \cite{barber_muse_2023}. It becomes only natural to consider the implications of such technologies when applied in the context of audio and music production.

Several generative models have been proposed for neural audio synthesis. NSynth \cite{engel_neural_2017} uses a WaveNet autoencoder to synthesize pitched instrument samples. GANSynth \cite{engel_gansynth_2019} considers an instantaneous frequency representation to model signal phase. \gls{ddsp} \cite{engel_ddsp_2020} and related works \cite{wu_sawsing_2022, nercessian_diffworld_2023} construct autoencoders with differentiable synthesizer back-ends to promote controllability. A real-time variational autoencoder design was introduced in \cite{caillon_rave_2021}. GANstrument \cite{narita_ganstrument_2023} utilizes a feature descriptor achieved through adversarial domain confusion. These models all lack an interface for controlling audio generation via text input. Consequently, we have witnessed a surge in the development of text-to-audio systems generating compelling audio examples from text prompts. One particular family of approaches rely on neural audio codecs \cite{zeghidour_soundstream_2021, kumar_high-fidelity_2023} representing audio compactly as a set of discrete codes whose sequence can be learned using transformer-based language models. While initial approaches targeted speech \cite{borsos_audiolm_2023, wang_valle_2023} and environmental sounds \cite{kreuk_audiogen_2023}, follow-on works adapt techniques for text-to-music, generating entire musical passages from text \cite{agostinelli_musiclm_2023, copet_simple_2023}.

In this paper, we introduce a new task which we call \emph{text-to-instrument}, whose aim is to generate musical instruments given a text prompt in a zero-shot manner. Under this task, we explicitly model a musical instrument as a collection of waveforms sampling an instrument's time-domain response across the axes of pitch (the fundamental frequency of a note) and velocity (the intensity with which a note is played). In this paradigm, we move beyond the constraints of any single parametric synthesizer, avoiding the expressivity limitations tied to its specific implementation details. As in \cite{narita_ganstrument_2023}, we note that injecting prior domain knowledge into the generative process via techniques like \glsentrytext{ddsp} is indeed interesting, but is complementary to this work as such approaches naturally constrain the manifold that system outputs can live on \cite{nercessian_diffworld_2023}. Unlike text-to-music, which predominantly involves generation of a single audio example for the text prompt at inference, text-to-instrument systems must generate an ensemble of audio examples such that they are individually stabilized in pitch and timbrally consistent with one another, so that they can be assembled into a playable instrument that can be triggered in predictable ways. In summary, our primary contributions are:

\begin{itemize}
    \item We introduce the text-to-instrument task.
    \item We propose \emph{InstrumentGen}, a catered text-to-instrument solution expanding on a state-of-the-art text-prompted generative audio model to be, alongside a \gls{clap} embedding \cite{wu_large-scale_2023} used as a joint audio/text representation, additionally conditioned on instrument family, source type (i.e. acoustic, electronic, or synthetic), pitch across the entire 88-key range of a standard full-length piano keyboard, and velocity.
    \item We present a differentiable loss to objectively assess the intra-instrument \gls{tc} of sample-based instruments for our task by generalizing a (potentially multi-scale) log mel spectrogram loss \cite{engel_ddsp_2020, berger_ddspeech_2020}, and use it as an evaluation metric in this work.
\end{itemize}

\section{Proposed Method}
\begin{figure}[b] 
\vspace{-1em}
\centering
\resizebox{1.0\textwidth}{!}{%

  \begin{tikzpicture}[
    encoderblock/.style={scale=1, shape=trapezium, trapezium angle=75, draw, shape border rotate=270, trapezium stretches=true, minimum width=20mm, minimum height=30mm, draw = color1, fill = color1!10, rounded corners= 3pt, line width = 1pt},
    decoderblock/.style={scale=1, shape=trapezium, trapezium angle=75, draw, shape border rotate=90, trapezium stretches=true, minimum width=20mm, minimum height=30mm, draw = color1, fill = color1!10, rounded corners= 3pt, line width = 1pt},
    block/.style={scale=1, shape=rectangle, draw, minimum width=15mm, minimum height=10mm, rounded corners= 3pt, line width = 1pt, draw = color3, fill = color3!10},
    linear/.style={scale=1, shape=rectangle, draw, minimum width=10mm, minimum height=5mm, rounded corners= 3pt, line width = 1pt, draw = color4, fill = color4!10},
    dot/.style={circle, fill,inner sep=1pt},
    loss/.style={scale=1, shape=rectangle, draw, minimum width=15mm, rounded corners= 3pt, line width = 1pt, draw = color2, fill = color2!10},
    font=\sffamily\small,
    text=black!80,
    ]
    \def\dx{5mm}
    \def\dy{10mm}
 
    \node (base) at (0mm, 0mm) {};

    \node (dac_encoder) [encoderblock, left=0mm of base, dashed] {\glsentrytext{dac} encoder};

    \node (clap_text_encoder) [encoderblock, below=25mm of dac_encoder, dashed, align=center, opacity=0.75, text opacity=1.0] {\glsentrytext{clap} (text)};
    
    \node (clap_encoder) [encoderblock, below=\dy of dac_encoder, dashed, align=center, opacity=0.75, text opacity=1.0] {\glsentrytext{clap} (audio)};

    \node (clap_text_encoder_input) [left=\dy of clap_text_encoder, align=center] {text prompt \\ (during inference)};

    \draw [-stealth] (clap_text_encoder_input) -- (clap_text_encoder);

    \coordinate (midpoint_dac_clap) at ($(dac_encoder.west)!0.5!(clap_encoder.west)$);
    
    \node [rotate=90, left=2cm of midpoint_dac_clap, anchor=center] (waveform_input_dac) {\includegraphics[width=4cm, height=1cm]{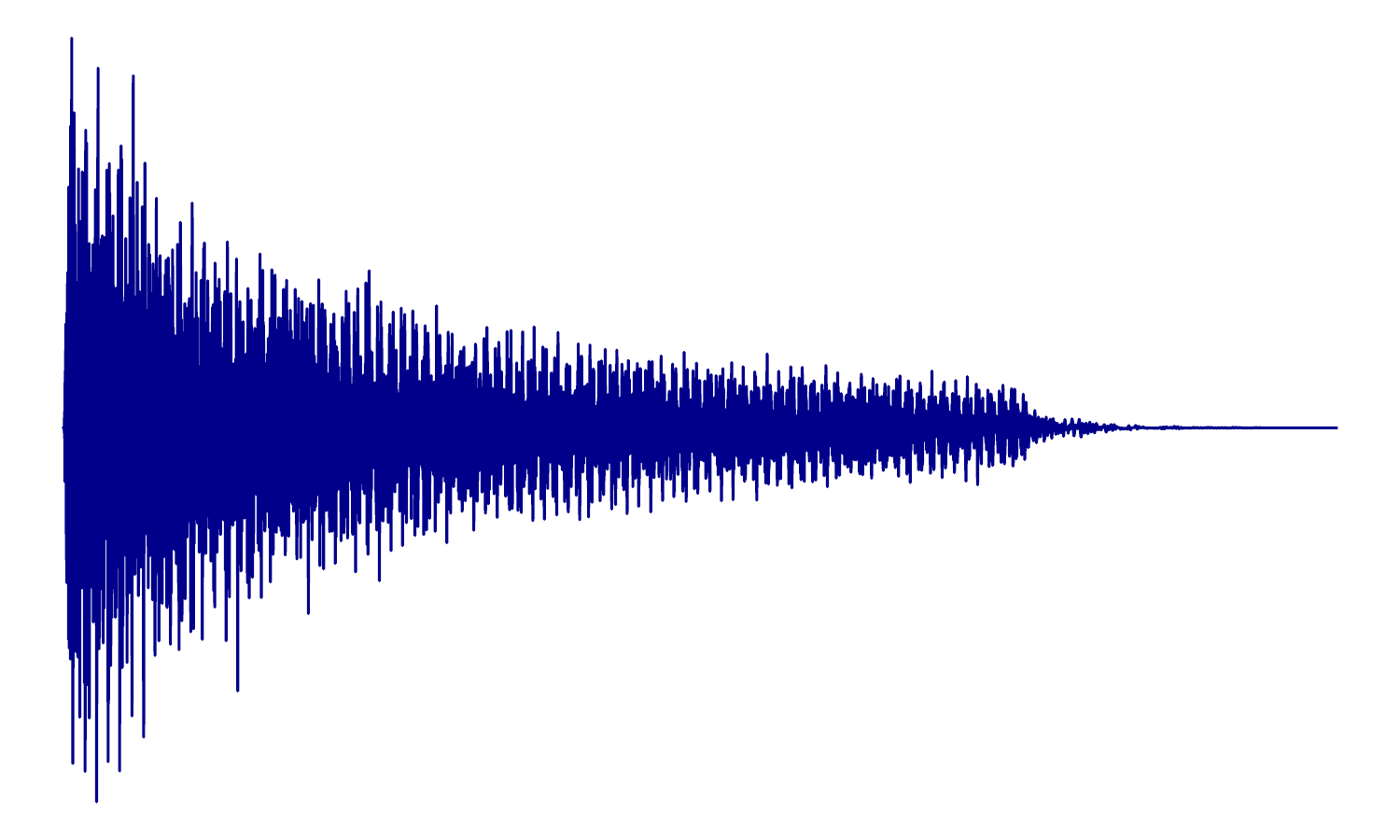}};
    %\node[rotate=90, anchor=north, align=center] at ($(waveform_input_dac.north) + (-5mm, 0)$) (waveform_input_dac_text) {single-shot sample \\ (only during training)};

    \node[rotate=90, anchor=north, align=center] at ($(waveform_input_dac.north) + (-5mm, 0)$) (waveform_input_dac_text) {single-shot sample \\ (during training)};

    \coordinate (midpoint_dac_clap_waveform) at ($(waveform_input_dac.south)!0.5!(midpoint_dac_clap)$);

    \draw [-stealth, rounded corners] (waveform_input_dac.south) -- (midpoint_dac_clap_waveform) node[near start, above] {$\mathbf{x}$} |-  (dac_encoder);
    \draw [-stealth, rounded corners] (waveform_input_dac.south) -- (midpoint_dac_clap_waveform) |-  (clap_encoder);

    \node (rvq_clap) [block, right=\dx of clap_encoder] {RVQ};

    \draw [-stealth, rounded corners] (clap_text_encoder) -| (rvq_clap);

    \node (rvq_linear) [linear, right=25mm of rvq_clap]{Linear};
    \node (rvq) [block, right=\dx of dac_encoder, dashed] {RVQ};
    \node (pattern) [block, right=\dx of rvq, align=center] {Codebook \\interleaving pattern};
         
    \node (lm) [block, right=12mm of pattern] {Transformer};
    \node (pattern2) [block, right=12mm of lm, align=center] {Codebook \\de-interleaving pattern};
    \node (dac_decoder) [decoderblock, right=\dx of pattern2, dashed] {\glsentrytext{dac} decoder};

    \coordinate (outpoint) at (midpoint_dac_clap -| dac_decoder.east);

    \node (waveform_output_dac) [rotate=90, right=2cm of outpoint, anchor=center] {\includegraphics[width=4cm, height=1cm]{waveform.pdf}};
    \node[rotate=90, anchor=north] at (waveform_output_dac.north) (waveform_output_dac_text) {reconstruction};

    %\node (waveform_output_dac) [rotate=90, xscale=0.8, yscale=0.4, right=2cm of outpoint, anchor=center] {\includegraphics[width=5cm]{waveform.pdf}};
    \draw [-stealth] (dac_decoder.east) -- ($(dac_decoder.east -| waveform_output_dac.north)$) node[midway, below] {$\mathbf{\hat{x}}$};

    \node (p_linear) [linear, below=5mm of rvq_linear]{LUT $\shortrightarrow$ Linear};
    \node (p) [left=10mm of p_linear]{pitch $\theta_\mathrm{p}$};
    
    \node (v_linear) [linear, below=1mm of p_linear]{LUT $\shortrightarrow$ Linear};
    \node (v) [left=10mm of v_linear]{velocity $\theta_\mathrm{v}$};
    
    \node (c_linear) [linear, right=11mm of v_linear]{LUT $\shortrightarrow$ Linear};
    \node (c) [right=10mm of c_linear]{instrument family $\theta_\mathrm{i}$};
    
    \node (s_linear) [linear, right=11mm of p_linear]{LUT $\shortrightarrow$ Linear};
    \node (s) [right=10mm of s_linear]{source type $\theta_\mathrm{t}$};

    \node (cat) [linear, right=\dx of rvq_linear] {$\mathrm{cat}(\cdot)$};

    \draw [-stealth] (dac_encoder) -- (rvq);
    \draw [-stealth] (rvq) -- (pattern) node[midway, below] {$\mathbf{c}$};
    \draw [-stealth] (pattern) -- (lm);
    \draw [-stealth] (lm) -- (pattern2);
    \draw [-stealth] (pattern2) -- (dac_decoder) node[midway, below] {$\mathbf{\hat{c}}$};
    \draw [-stealth] (clap_encoder) -- (rvq_clap);
    \draw [-stealth] (rvq_clap) -- (rvq_linear) node[midway, above] {$\bm{\theta}_{\mathrm{clap}}$};
    \draw [-stealth] (rvq_linear) -- (cat);

    \draw [-stealth] (p) -- (p_linear);
    \draw [-stealth] (v) -- (v_linear);
    \draw [-stealth, dotted] (c) -- (c_linear);
    \draw [-stealth, dotted] (s) -- (s_linear);
    
    \draw [-stealth, rounded corners] (p_linear) -| ($(cat.south) + (-0.3cm, 0cm)$);
    \draw [-stealth, rounded corners] (v_linear) -| ($(cat.south) + (-0.1cm, 0cm)$);
    \draw [-stealth, rounded corners] (c_linear) -| ($(cat.south) + (0.1cm, 0cm)$);
    \draw [-stealth, rounded corners] (s_linear) -| ($(cat.south) + (0.3cm, 0cm)$);
    
    \draw [-stealth, rounded corners] (cat) -| (lm.south) node[near start, above] {$\bm{\theta}$} node[near end, right] {cross attention};

    % Loss function box
    %\coordinate (midpoint) at ($(pattern.east)!0.5!(lm.east)$);
    \node[loss, above=5mm of lm] (loss_box) {\(\mathcal{L}_\mathrm{ce}\)};

    \node[dot] (dot1) at ($(rvq.east)!0.5!(pattern.west)$) {};
    \node[dot] (dot2) at ($(pattern2.east)!0.5!(dac_decoder.west)$) {};

    \draw [-stealth, dashed, rounded corners] (dot1) |- (loss_box.west);
    \draw [-stealth, dashed, rounded corners] (dot2) |-  (loss_box.east);

  \end{tikzpicture}
  }
  \caption[]{Overview of proposed method. Dashed lines indicate modules with fixed and pretrained weights during training. Source type and instrument family are not provided during inference.}
  \label{fig:model_overview}
\end{figure}

InstrumentGen is based on the MusicGen \cite{copet_simple_2023} architecture as a foundation, which consists of a neural audio codec and a language model designed to predict acoustic tokens based on conditioning signals. To improve audio quality, we replace the original EnCodec architecture \cite{defossez_high_2022} used in MusicGen with the \gls{dac} \cite{kumar_high-fidelity_2023}, which addresses the issue of codebook collapse in previous models while achieving higher audio fidelity. Additionally, we introduce a set of new conditioning signals to the system: this includes instrument family, source type, pitch, and velocity, alongside a joint language-audio embedding \cite{wu_large-scale_2023}.  This conceivably allows instrument samples to be inferred from either text or audio prompts, where we focus on the former but can also perform the latter.
Fig.~\ref{fig:model_overview} gives an overview of our method.

\subsection{Compressed Audio Representation}
In this work, we employ the \glsentrytext{dac} as an intermediate representation of the monophonic input waveform $\mathbf{x} \in \mathbb{R}^{1 \times L}$ (cf. Fig.~\ref{fig:model_overview}), resulting in the discrete codes $\mathbf{c} \in \mathbb{R}^{C \times N}$. Here, $L$ denotes the length of the waveform, $N$ the sequence length of the acoustic tokens, and $C$ the number of codebooks used.
The \glsentrytext{dac} is trained on a broad spectrum of audio types, thereby making it also suitable for generating tonal one-shot instrumental sounds. We deliberately opt to model our task at a sampling rate of 44.1 kHz, as this would ultimately be a minimum requirement for real-world music production applications. We employ the corresponding pretrained model weights and fix them during training.

\subsection{Language Model}

For modeling the discrete audio tokens of single-shot instrumental samples, we consider a smaller, 60M parameter variant of the MusicGen transformer decoder \cite{copet_simple_2023}, both to prevent overfitting and to provide faster inference. The resulting model consists of 12 decoder layers with 16 attention heads per layer and a transformer model dimension of 512. As in MusicGen \cite{copet_simple_2023}, we predict and reconstruct audio from tokens of the 4 most significant \cite{kumar_high-fidelity_2023} codebooks at each frame. Our predictor model learns to select tokens from codebooks of size 1024 using delayed pattern interleaving \cite{copet_simple_2023}.

\subsection{Conditioning}
\paragraph{Categorical conditioning} 
We use a categorical conditioning scheme for instrument family $\theta_{\mathrm{i}}$, source type $\theta_{\mathrm{t}}$, pitch $\theta_{\mathrm{p}}$, and velocity $\theta_{\mathrm{v}}$, that consists of a lookup table (LUT) and a fully connected layer that maps the dimension of the categorical feature space to the inner dimension of the language model. When used, the instrument family and source type attributes in our dataset serve as supplementary metadata-driven timbral cues for generation. For pitch, we model the $P=88$ range of notes spanned by a standard full-length keyboard, corresponding to \gls{midi} note numbers 21-108, and note that this is a significant expansion relative to the chroma feature used in \cite{copet_simple_2023}. We consider $V=5$ velocity layers, according to the values that are actually present within our training dataset, corresponding to \glsentrytext{midi} velocities 25, 50, 75, 100, and 127. 

\paragraph{Joint text and audio conditioning} 
Wu et al. \cite{wu_large-scale_2023} introduced a \glsentrytext{clap} framework, which is designed to process both audio and text data and generate corresponding embeddings by using two separate encoders: one for audio and another for text. Both embeddings are further processed by a 2-layer MLP with ReLU activation to bring them into a fixed dimension of $512$. The resulting model leverages a contrastive loss function, and is trained on musical signals to encourage the audio and text embeddings to be similar when they come from matching pairs. The audio encoder uses the HTS-AT (transformer-based) architecture \cite{chen_hts-at_2022}, while the text encoder leverages RoBERTa \cite{liu_roberta_2019}.

When \glsentrytext{clap} embeddings are used, we quantize them via \gls{rvq} with learned codes, yielding $\bm{\theta}_{\mathrm{clap}}$. 
 Since the various conditioning signals used here reflect global cues $\bm{\theta}$ for steering audio generation, they are fused with the transformer decoder by means of cross-attention.
\begin{figure*}[b]
    \vspace{-2em}
  \centering
  \subfloat[]{
    \includegraphics[width=0.24\textwidth]{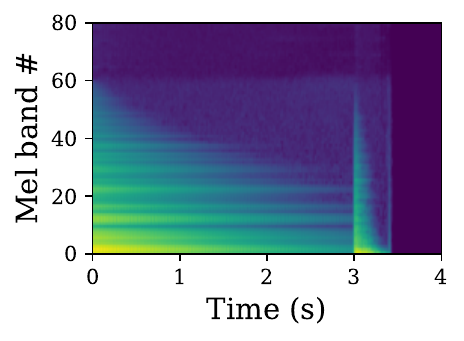}
    \includegraphics[width=0.24\textwidth]{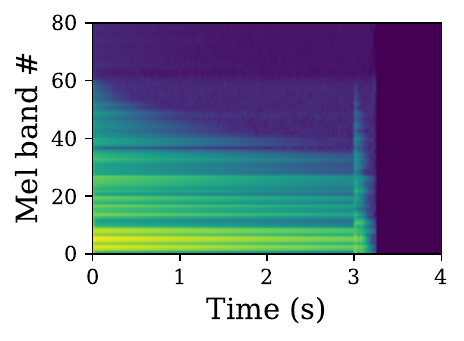}
  }\hfill
  \subfloat[]{
    \includegraphics[width=0.24\textwidth]{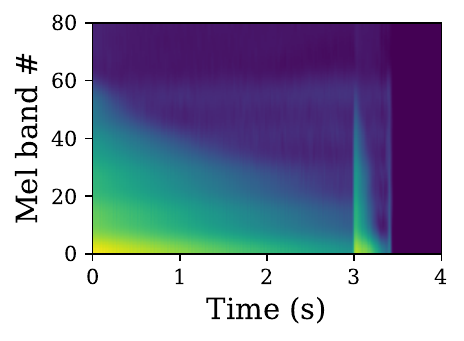}
    \includegraphics[width=0.24\textwidth]{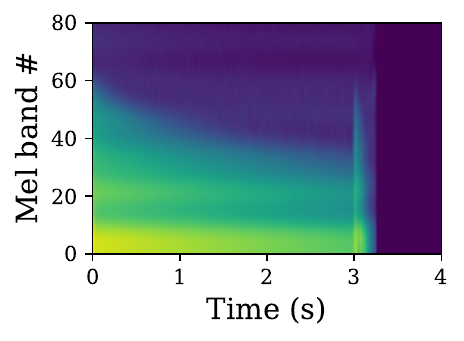}
  }
  \caption{Log mel spectrograms of \texttt{guitar$\_$acoustic$\_$010-024-100} (C1, left) and \texttt{guitar$\_$acoustic$\_$010-049-100} (C\#3, right) (a) without and (b) with liftering. Liftering can allow examples of differing pitch to be more readily compared to one another in the timbral sense.}
  \label{fig:timbre}
\end{figure*}

\section{Timbral Consistency Measure}
Our task necessitates that waveforms comprising a generated instrument are timbrally consistent to one another. We provide an objective means to measure this relation, considering that signals within an instrument may vary in pitch. %, which can complicate their direct comparison timbrally. 
 Specifically, \gls{mfccs} are ubiquitously known as timbral descriptors, as a cepstral lifter whose quefrencies correspond to the first handful of \glsentrytext{mfccs} can be used to estimate a signal's spectral envelope \cite{terasawa_timbre_2006}. We integrate this notion into a differentiable objective function which generalizes a multi-scale log mel spectrogram loss \cite{berger_ddspeech_2020}, aggregating over all possible pairwise comparisons within the instrument. For an ensemble of waveforms $\mathbf{X} \in \mathbb{R}^{K \times L}$ comprising an instrument (with $K = PV$), the \glsentrytext{tc} measure is defined as
\begin{equation}
\mathcal{L}_{\glsentrytext{tc}}(\mathbf{X}) = \sum\limits_{i=1}^{K}\sum\limits_{j=i+1}^{K}\sum\limits_{s=1}^{S}\|\mathbf{y}_i - \mathbf{y}_j \|^1_1
\end{equation}
over $S$ scales, where for an arbitrary audio waveform $\mathbf{x}_k$ from $\mathbf{X}$ we have 
\begin{equation}
\mathbf{y}_k = \mathbf{D}_{M_s}^{-1}\mathbf{D}_{m_s}\log\left[ \mathbf{B}_{s}\left|\mathcal{F}_s(\mathbf{x}_k)\right|^p\right]
\end{equation}
For each scale $s$, $\mathcal{F}_s$ denotes the respective short-time Fourier transform, $\mathbf{B}_s$ denotes an $M_s$-band mel transformation matrix, and $\mathbf{D}_{m_s} \in \mathbb{R}^{M_s \times M_s}$ denotes a discrete cosine transform basis matrix that has been masked with zeros for row indices greater than $m_s$ \glsentrytext{mfccs}. When $m_s=M_s$ (i.e. all available \glsentrytext{mfccs} are considered), $\mathbf{D}_{M_s}^{-1}\mathbf{D}_{M_s}=\mathbf{I}$. With $m_s < M_s$, we effectively apply cepstral liftering that can neutralize the effect of the pitched excitation within the signal (cf. Fig.~\ \ref{fig:timbre}). We consider $S=1$, $M_s=80$, and $m_s=13$, and $p=1$ for demonstrative purposes in this work.

\section{Experimental Results}

We train models on the NSynth dataset \cite{engel_neural_2017}, pruning it according to our specified 88-key pitch range. We resample the dataset, captured at 16~kHz, to a target rate of 44.1~kHz, viewing it as a proxy in lieu of an equally comprehensive full-band alternative. Models are trained to minimize the cross-entropy $\mathcal{L}_\mathrm{ce}$ between predicted codes $\mathbf{\hat{c}}$ and ground truth $\mathbf{c}$, over 1M training steps with AdamW optimizer, a batch size of 32, and a cosine-annealed schedule with initial learning rate of $10^{-3}$ as in \cite{copet_simple_2023}.

We consider two models to evaluate the fundamental capabilities of the system under investigation. In the first model, instrument timbre is specified from a closed set of instrument family and source type fields available as metadata in the dataset. Our second model is truly a text-to-instrument model, where we leverage \glsentrytext{clap} to enable text input at inference. In this case, the instrument family and source type attributes help guide training (subject to dropout with 70\% probability), but do not need to be specified at inference. Pitch and velocity are expected as inputs during training and inference across all models in order to generate a complete sample-based instrument. We explicitly note that we did not condition models on more specific instrument metadata because training and evaluation datasets constitute disjoint sets of instruments, and cannot compare to existing text-to-music systems because they are constrained in their specificity of pitch and velocity needed to carry out our task.

We compare models quantitatively (cf. Tab.~\ref{tab:quant}), generating instrument samples reflecting those in the NSynth test set according to a fixed input cue representing each instrument (i.e. a single metadata configuration or \glsentrytext{clap} embedding). As is customary \cite{agostinelli_musiclm_2023, copet_simple_2023}, we report \gls{fad} leveraging VGGish and average \glsentrytext{clap} score. To measure the efficacy of our pitch conditioning, we estimate the median pitch of generated samples using YIN \cite{cheveigne_yin_2002}, reporting \gls{mad} relative to ground truth. Moreover, we measure the average \glsentrytext{tc} across all instruments. Despite potential estimation errors, the detected pitches of our outputs are generally well within a semitone of their targets. The \glsentrytext{clap}-based model noticeably improves upon the simple metadata-driven model in terms of \glsentrytext{clap} and \glsentrytext{tc}. Its \glsentrytext{tc} metric approaches that of ground truth, with some gap left to close in future works. Interestingly, the metadata-driven model yields marginally lower \glsentrytext{fad}, which may be due to VGGish not being specifically fine-tuned to musical audio signals. 

Lastly, we curate a set of text prompts, generating corresponding instruments using our \glsentrytext{clap}-based system. We report average \glsentrytext{clap} score correlating generated instruments to their text prompts, \glsentrytext{mad}, and \glsentrytext{tc} (cf. Tab. \ref{tab:qual}). We compile 1-5 scale \gls{mos} across members of our organization for quality (\glsentrytext{mos}$_\mathrm{quality}$), text correspondence (\glsentrytext{mos}$_\mathrm{text}$), and \glsentrytext{tc} (\glsentrytext{mos}$_\mathrm{\glsentrytext{tc}}$). We refer readers to our supplementary materials, available at \url{https://instrumentgen.netlify.app}.
\section{Conclusion}
In this work, we introduced \emph{text-to-instrument}, and proposed a neural audio codec language model that is catered for the task. We highlighted the fundamental difference between text-to-instrument and other related tasks, whereby the former must generate several samples corresponding to the text prompt that are timbrally consistent to one another. Moreover, we suggested a differentiable objective for measuring the timbral consistency of generated musical instruments. We established a baseline which can generate timbrally consistent sample-based instruments, where we have enabled disentanglement of pitch and timbre via the cross-attention of various conditioning signals incorporated within our system. Future work will consider additional techniques to further improve the fidelity of our system.
\begin{table}[h]
\vspace{-1.0em}
\centering
  \begin{tabular}{lcccc}
    \toprule
     Model & \glsentrytext{fad}$\downarrow$ & \glsentrytext{clap}$\uparrow$ & \glsentrytext{mad}$\downarrow$ & \glsentrytext{tc}$\downarrow$ \\ \midrule
     Instrument family/source type & \textbf{1.631} & 0.487 & \textbf{0.045} & 1.813 \\
     \glsentrytext{clap} & 1.692 & \textbf{0.691} & \textbf{0.045} & \textbf{1.236} \\ \midrule
     Ground truth & -- & -- & -- & \underline{\textbf{1.053}} \\ \bottomrule
  \end{tabular}
 \caption{Evaluation over the NSynth test set.}
 \label{tab:quant}
\end{table}
\begin{table}[h]
\vspace{-2.0em}
\centering
  \begin{tabular}{lcccccc}
    \toprule
     Model & \glsentrytext{clap}$\uparrow$ & \glsentrytext{mad}$\downarrow$ & \glsentrytext{tc}$\downarrow$ & \glsentrytext{mos}$_\mathrm{quality}\uparrow$ & \glsentrytext{mos}$_\mathrm{text}\uparrow$ & 
     \glsentrytext{mos}$_\mathrm{\glsentrytext{tc}}\uparrow$ \\ \midrule
     \glsentrytext{clap} & 0.235 & 0.162 & 0.873 & 3.094 & 3.620 & 3.465 \\
     \bottomrule
  \end{tabular}
 \caption{Evaluation over a curated set of text prompts.}
 \label{tab:qual}
 \vspace{-1em}
\end{table}
\bibliography{neurips_2023}
\bibliographystyle{ieeetr}
\setcitestyle{square}
\end{document}